\documentclass[twocolumn,secnumarabic,amssymb, nobibnotes, aps, prd, superscriptaddress]{revtex4}

\usepackage{graphicx}% Include figure files
\usepackage{dcolumn}% Align table columns on decimal point
\usepackage{bm}% bold math

\def\beq{\begin{eqnarray}}
\def\eeq{\end{eqnarray}}
\def\e{{\rm e}}

\begin{document}

% Use the \preprint command to place your local institutional report
% number in the upper righthand corner of the title page in preprint =
% mode.
% Multiple \preprint commands are allowed.
% Use the 'preprintnumbers' class option to override journal defaults
% to display numbers if necessary
%\preprint{}

%Title of paper
  \title{Lorentz symmetric dispersion relation from a random Hamiltonian}

% repeat the \author .. \affiliation  etc. as needed
% \email, \thanks, \homepage, \altaffiliation all apply to the current
% author. Explanatory text should go in the []'s, actual e-mail
% address or url should go in the {}'s for \email and \homepage.
% Please use the appropriate macro foreach each type of information

% \affiliation command applies to all authors since the last
% \affiliation command. The \affiliation command should follow the
% other information
% \affiliation can be followed by \email, \homepage, \thanks as well.
\author{Andreas Albrecht}
%\email[]{Your e-mail address}
%\homepage[]{Your web page}
%\thanks{}
%\altaffiliation{}
%% \affiliation{University of California at Davis\\ Department of
%%   Physics\\ One Shields Avenue \\ Davis, CA 95616 USA}
\author{Alberto Iglesias}
\affiliation{University of California at Davis\\ Department of
  Physics\\ One Shields Avenue \\ Davis, CA 95616 USA}
%Collaboration name if desired (requires use of superscriptaddress
%option in \documentclass). \noaffiliation is required (may also be
%used with the \author command).
%\collaboration can be followed by \email, \homepage, \thanks as well.
%\collaboration{}
%\noaffiliation

%\date{\today}

\begin{abstract}
We match the density of energy eigenstates of a local field theory
with that of a random Hamiltonian order by order in a Taylor
expansion.  In our previous work we assumed Lorentz symmetry of the field
theory, which entered through the dispersion relation. Here we extend
that work to consider 
a generalized dispersion relation and show that the Lorentz symmetric
case is preferred, in that the Lorentz symmetric dispersion relation
gives a better approximation to a random Hamiltonian than the other
local dispersion relations we considered. 
\end{abstract}

% insert suggested PACS numbers in braces on next line
% insert suggested keywords - APS authors don't need to do this
%\keywords{}

%\maketitle must follow title, authors, abstract, \pacs, and \keywords
\maketitle

% body of paper here - Use proper section commands
% References should be done using the \cite, \ref, and \label commands
\section{Introduction}
\label{Sect:I}
We consider the question ``can a random Hamiltonian
be interpreted as a field theory to a sufficiently good
approximation to describe the observed physical
world?''.  We posed this question in earlier work
\cite{Albrecht:2007mm,Albrecht:2008bj} from the point of view of a
comparison of the energy eigenvalue spectrum, and found an
affirmative answer.
In that work we considered only field theories with a dispersion
relation given by Lorentz symmetry.  The dispersion 
relation is the only
way the presumed Lorentz invariance of the field theory entered into
our calculations. In this
work we extend our previous analysis to include a one parameter family
of possible dispersion relations. We also include the dimension of space $d$
(as was already analyzed in \cite{Albrecht:2008bj}).  Our analysis
shows that choosing a Lorentz invariant dispersion relation makes it
easier to approximate the field theory density of states by that of
a random Hamiltonian.  We thus conclude that in a physical picture where
local field theory is just an approximate interpretation of a physical
world which is fundamentally described by a random Hamiltonian,
field theories with a Lorentz invariant dispersion relation are
favored.

As emphasized in our earlier work, the comparison at the level of the
density of energy eigenstates is only one check one can make on this
set of ideas.  To more fully develop our line of thinking one must
explore the mapping between energy eigenstates and the fields representing
observable particles or their constituents. This analysis could lead in
principle  to some symmetries and representations being favored
over others.  If such predictions are sharp enough one may even have a
chance to falsify these ideas. In this paper we limit ourselves to a
comparison at the level of energy spectra.  We find it intriguing that
already at this level the Lorentz invariant dispersion relation is
favored. 
Still, our results are only 
sensitive to Lorentz invariance through the dispersion relation, therefore, 
our approach is unable to discern a Lorentz invariant theory from a 
non-invariant one that has the same dispersion relation -- such as 
field theories in noncommutative spacetime \cite{Chaichian:2004za}
which generically break Lorentz symmetry but preserve $P^2$,
unaltered, as a Casimir operator.  
To discriminate such theories other quantities 
beyond the densities of states should be probed involving other quantum 
numbers such as spin, for example, to specify the field content and the 
invariances unambiguously. 

Our past work on this topic was motivated by the 
``clock ambiguity'' in quantum gravity. That motivation and
how it takes us to the comparison of energy spectra is discussed at
length in \cite{Albrecht:2007mm} and \cite{Albrecht:2008bj}.   In this
  paper we focus  
more narrowly on extending our formalism to generalized
dispersion relations and interpreting our results.  We feel that this
technical work will be of interest to a wider group that may bring
different motivations than our own.   In section
\ref{Sect:Approach} we lay out our basic technical approach in a manner that
is fairly self-contained.  The
key new technical ingredient of this paper is our derivation of the
density of states for field theories with a generalized dispersion
relation. We present two such derivations, one using statistical 
mechanics (section \ref{Sect:Stat}) and one using scaling arguments
(section \ref{Sect:Scale}), and discuss the relationship between these
two approaches in section \ref{Sect:Compare}. In section
\ref{Sect:Analysis} we integrate the generalized dispersion relation
results into the framework presented in section \ref{Sect:Approach}
and show how this leads to our main conclusions.  Section
\ref{Sect:Conclusions} reviews our conclusions and explores some
interesting directions for further study.

\section{Our basic approach}
\label{Sect:Approach}

If the evolution of the world around us were most fundamentally
described by a random Hamiltonian, could that Hamiltonian be
approximated to a sufficient degree by a field theory to allow us to
{\em interpret} the physical world in terms of a field theory?
In \cite{Albrecht:2007mm} we addressed that question at the level of
the spectrum of energy eigenvalues.  We found that the spectrum of a sufficiently large random
Hamiltonian could approximate that of a free field theory over a
sufficiently large range of energies to allow the former to be
approximated by the latter on scales relevant to our data about the
physical world.  In this section we review that work with an eye
toward the topic of this paper: The extension of that analysis to
field theories with a generalized dispersion relation. 

The distribution of eigenvalues for a random Hamiltonian, represented as an
$N_H\times N_H$ Hermitian matrix, follows under quite general assumptions %\cite{Mehta} 
the Wigner semicircle rule in the large $N_H$ limit, for example, the
distribution of eigenvalues of a large Hermitian matrix with elements
drawn from a Gaussian distribution.

On the other hand, the density of states for a free field
theory grows at large energies like an exponential of a 
power of the energy.  On the face of it, these two forms for $dN/dE$
are dramatically different. In order to press forward with the comparison we 
introduced a general parametrization for the random Hamiltonian and field theory
spectral densities respectively:
\begin{eqnarray}\label{dr}
{dN_R\over dE}&=&\!\!\!\left\{
\begin{array}{ll}
a{N_H\over E_M}\left(1-\left({E-E_S\over E_M}\right)^\beta\right)^\gamma & 
|E-E_S|<E_M
\\
0 & {\rm otherwise}~, \end{array}
\right.\\
{dN_F\over dE}&=&{B\over E}\exp\left\{b\left(E\over
k_V\right)^\alpha\right\}~.\label{df}
\end{eqnarray}
The parameters  $E_M$ and $E_S$ represent the maximum eigenvalue of the random
Hamiltonian and an offset energy between the two descriptions, $k_V$
($\equiv 2\pi/L$) is the resolution in $k$-space set by putting the field theory in a box of size $L$ 
and $B$, $b$, $\alpha$ and $\gamma$ are dimensionless parameters.
Since we have not yet explored the emergence of gravity in this picture
we initially allow the energy offset $E_S$ to be a free parameter.  The
value of $k_V$ sets a scale at which continuum field theory
breaks down. 
The standard expression
\begin{equation}
  N = \exp{S}
\label{NofS}
\end{equation}
leads to 
\begin{equation}
B = \alpha S~.
\label{BofS}
\end{equation}
Which relates parameters in Eqn. \ref{df} to the entropy $S$.
Throughout this paper we we use units where $\hbar = c = k_B = 1$.

We expand both Eqns.~\ref{dr} and \ref{df} in a Taylor series
around a given central energy $E_0=\rho_U V_U=10^{80}GeV$ (using the
energy density $\rho_U$ and volume $V_U$ of the observed Universe today).
We explore the discrepancies at each order in $(E-E_0)$ to find the level of agreement between the two descriptions.

Requiring exact equality at zeroth order sets the size of the space of the random Hamiltonian 
to be exponentially large:
\beq\label{m1}
N_H = {B\over a}{E_M\over E_0}
\left[1-\left(E_0-E_S\over E_M\right)^\beta\right]^{-\gamma}
\!\!\!\exp \left[b\left(E_0\over k_V\right)^\alpha\right]~.
\eeq
More generally, this expression should be seen to only give a lower bound on $N_H$, since we 
currently only know upper bounds on $k_V$ (which gives a scale of breakdown of
continuum field theory).  

\begin{figure}
\includegraphics[width=3.3in]{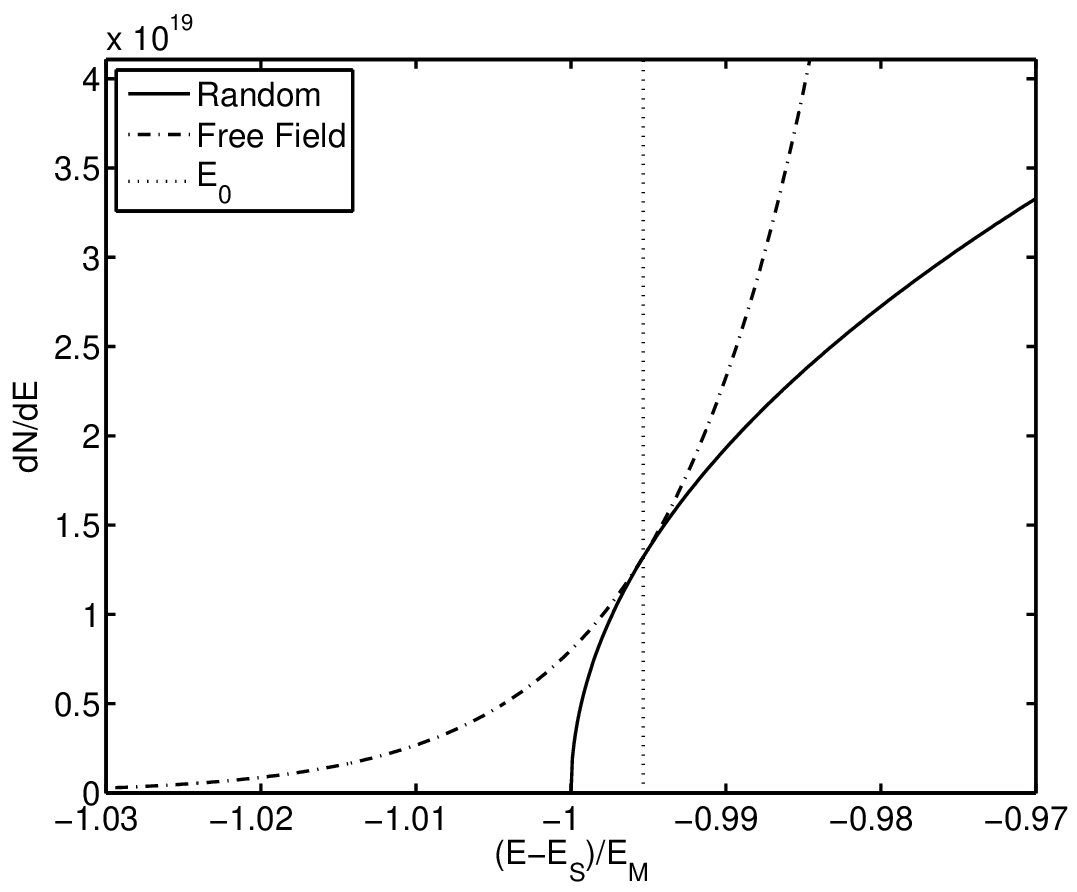}
\caption{\label{fig:both}
A plot of the density of energy eigenvalues for a random Hamiltonian using
Eqn.~\ref{dr} and a field theory using Eqn.~\ref{df} matching the zeroth and 
first order terms in a Taylor expansion around $E_0$ (the vertical line).  
}
\end{figure}

Requiring equality at first order (as well as equality at zeroth order) sets the offset 
energy $E_S$ in terms of the energy of the Universe $E_0$ by the
following implicit expression:
\beq\label{m2}
-\beta\gamma{E_0\over E_0-E_S}
{\left(E_0-E_S\over E_M\right)^\beta\over 1-\left(E_0-E_S\over E_M\right)^\beta}
=\alpha b \left(E_0\over k_V\right)^\alpha~.
\eeq
Figure \ref{fig:both} gives examples of the two density of states
curves equated at zeroth and first order according to the above analysis.
Assuming equality and 0th and 1st order, the relative difference at
second order is fixed and given by
\begin{eqnarray}\label{delta2}
\Delta_2\equiv {\Delta {dN\over dE}\over {dN\over dE}|_{E_0}}\approx
{\alpha^2 b^2\over\gamma}
\left(E_0\over k_V\right)^{2\alpha} {(E-E_0)^2\over E_0^2}~.
\end{eqnarray}
In \cite{Albrecht:2007mm} we considered specific values for the
parameters in Eqn. \ref{delta2} based on properties of the observed
physical world and found that $\Delta_2$ is small for realistic
values.  We concluded that the hypothesis that the laws of
physics could most fundamentally be given by a random Hamiltonian, and
just interpreted to a good approximation as a field theory had passed
our ``energy spectrum test''. 

The goal of this paper is to repeat the above analysis with a
generalized dispersion relation used to get the field theory density
of states.  Since Lorentz invariance enters only though the dispersion
relation, we hope to learn to what extent Lorentz invariant field
theories are preferred when a random Hamiltonian is approximated by a
field theory. 
To this end we introduce a general dispersion relation with two 
parameters, $m$ and $g$, of the following form: 
\begin{equation}
  {\cal E} \left( p \right) = p{\left( {\frac{p}{m}} \right)^g}~.
\label{gendisp}
\end{equation}
Here ${\cal E}$ represents the energy of a single particle with momentum $p$,
and should not be confused with $E$, the energy at which we evaluate
the density of states $N(E)$.  In general $E$ represents the energy
of multi-particle states (generally comprised of particles with
with more than one energy).
When the power $g$ vanishes we recover the dispersion relation of relativistic
massless particles and deviations from $g=0$ break Lorentz symmetry. The 
energy scale $m$ is needed to keep the dimensions correct for nonzero
values of $g$.   A discussion of possible values of $m$, which is
critical for the interpretation of this work, takes place in Section \ref{Sect:Analysis}.

There are a number of arguments to discard dispersion relations with negative 
values of $g$ (see \cite{Mattingly:2005re} and references therein for 
related studies); a very direct one is as follows:
If we allow $g<0$, the group velocity, $\partial{\cal  E}/\partial p$ 
(proportional to $p^g$), becomes divergently large for packets with 
components of low momentum. Thus, if we require signals to propagate at a
finite speed (below a certain cutoff of accessible momenta) we are
restricted to consider non-negative values of $g$.

We conclude this section with some comments about additional
assumptions and points of view implied by our framework.  The formulas
for the field theory
densities of states quoted here (and generalized in the rest of this
paper using Eqn. \ref{gendisp}) assume flat space (Minkowski spacetime for
g=0). Following the approach developed in earlier work
\cite{Albrecht:2007mm,Albrecht:2008bj} and similar in spirit to what
we have discussed here for field theory,  we consider the possibility that
general relativity gives only an approximate description of physics that need
only be valid around the observed state of the Universe.  Thus the
field theory in Minkowski space and its extensions which we consider
here, without any explicit account made for gravity, seem like a
reasonable place to start.  To the
extent that we can argue here that Lorentz invariant theories are
favored we will be led to consider fields in different
representations of the Lorentz group, including spin-2.  This can lead
to the emergence of general relativity, as discussed briefly in
section \ref{Sect:Conclusions}.

\section{The density of states for a generalized dispersion relation:
  Stat mech approach}
\label{Sect:Stat}

\subsection{Outline}
\label{SMout}

For a thermodynamic system at temperature $T$ in a volume $V$, 
the density of states is given by 
\begin{equation}
\frac{dN(E)}{dE} = \frac{d}{dE}\e^{S(E)}~,
\label{dndEofS}
\end{equation}
where
\begin{equation}
  S \equiv Vs = V\frac{\rho + P}{T}~,
\end{equation}
gives the (extensive) entropy $S$ in terms of the intensive quantities
pressure $P$ and energy density $\rho$.   In this section we construct
statistical mechanical expressions for $S(T)$ and $E(T)$ and then
invert the latter to allow us to write 
\begin{equation}
  S(E) = S(T(E))~.
\end{equation}
Although such a manipulation is only valid in the thermodynamic limit,
we expect this to be a good approximation for a system the size of the
observable Universe. 

\subsection{Energy density}

For a homogeneous statistical system of degrees of freedom of degeneracy 
$n$ in $d$ space dimensions the energy density is given by
\begin{equation}
\rho(T) = \frac{n}{(2\pi)^d}\int {\cal E}(p)f(p)d^dp~.
\label{rhoofT}
\end{equation}
Throughout this paper we take  degeneracy $n=1$ for simplicity, without
impacting our main points. For a free field theory in states of no net charge 
(and thus zero chemical potential, characteristic of the observed Universe) 
the momentum distribution $f(p)$ is given by 
\begin{equation}
  f(p) = \frac{1}{\e^{{\cal E}(p)/T} \pm 1}
\end{equation}
where the upper/lower signs are for Fermi/Bose statistics respectively. 

Using Eqn.~\ref{gendisp} in Eqn.~\ref{rhoofT} gives
\begin{eqnarray}
  \rho (T) &=& \frac{n}{(2\pi)^d}\int  
  \frac{p\left(\frac{p}{m}\right)^g}{\e^{p\left(\frac{p}{m}\right)^g/T}\pm 1} d^dp 
\nonumber\\
&=& T\left(Tm^g\right)^{d/\left(g+1\right)}F(g,d)~.
\label{rho}
\end{eqnarray}
Here $F$ is $O(1)$ and is given by
\begin{eqnarray}
  F(g,d) &\equiv& \frac{n}{(2\pi)^d}
  \int \frac{z^{g+1}}{\e^{z^{g+1}} \pm 1}d^dz \nonumber\\
   &=&\frac{n}{(2\pi)^d} \frac{2\pi^{d/2}}{\Gamma (d/2)}\frac{1}{g+1}
\int_0^\infty\frac{y^{d/(g+1)}}{\e^y\pm 1}dy~. 
\label{Fdef}
\end{eqnarray}
To arrive at Eqn.~\ref{Fdef} we changed variables in Eqn.~\ref{rho} 
using $z\equiv p\left(Tm^g\right)^{-1/(g+1)}$ 
and then $y\equiv z^{g + 1}$.  We show explicitly in
section \ref{Sect:Analysis} that $F(g,d)$ is sufficiently slowly 
varying and close to unity that it can be safely neglected when
building intuition about the expressions we are calculating.  (Still, for
completeness we keep it in for our final analysis.)

\subsection{Pressure}
There are some additional subtleties in defining pressure for a
generalized dispersion relation.
We start by rewriting the thermodynamic relation $P=-\partial E/\partial V$
in the following way 
\beq
P=-\frac{\partial E}{\partial V}=-\frac{n}{(2\pi)^d}\int 
\frac{\partial p}{\partial V} 
\frac{d{\cal E}(p)}{dp} f(p)d^dp
\eeq
Next, as in standard kinetic theory, we associate a volume expansion 
$\delta V=A\delta x$ to the pressure on the area $A$ due to 
elastic reflection of a particle with momentum $p_x$, that results in 
$\delta p_x=-2p_xA \delta x$. 
For a expansion in $d$ dimensions this generalizes to 
\beq
\frac{\partial p}{\partial V}=-\frac{2}{d}p~,
\eeq
since $\delta x\sim \delta r/\sqrt{d}$ and $p_x\sim p/\sqrt{d}$. 
Thus,
\beq
P=\frac{2}{d}\frac{n}{(2\pi)^d}\int p\frac{d{\cal E}}{dp} f(p)d^dp~,
\eeq  
Using the dispersion relation of Eqn.~\ref{gendisp}, namely, 
\beq
p \frac{d{\cal E}(p)}{dp}=(g+1){\cal E}(p)~,
\eeq
we find
\beq
P=\frac{2g+2}{d}\rho~.
\eeq

\subsection{The statistical result}
 We find it convenient to write the volume $V$ in terms of the inverse length 
given by
\begin{equation}
  k_V \equiv V^{-1/d}~.
\end{equation}
We are now in a position to complete the approach outlined in section
\ref{SMout} using the expressions for $P(T)$ and $\rho(T)$ derived in
the previous two sections.  The result is 
\begin{eqnarray}
&&S(E,g,d)=\nonumber\\
&&~\frac{d+2g+2}{d}F^{\frac{g+1}{d+g+1}}
\left(\frac{E}{k_V}\right)^{\frac{d}{d+g+1}}
\left(\frac{m}{k_V}\right)^{\frac{gd}{d+g+1}}.
\label{SEgd}
\end{eqnarray}
Note that both of the
exponentiated factors in Eqn. \ref{SEgd} are greater
than unity.  The quantity $k_V$ sets the energy scale for the breakdown of
field theory, and it should certainly be smaller than $m$ for the
generalized dispersion relation to have any meaning, leading to
$m/k_V>1$.  The factor with $E$ is discussed explicitly in section 
\ref{Sect:Approach}.

\section{The density of states for a generalized dispersion relation:
  Scaling approach}
\label{Sect:Scale}

\subsection {Outline}
In this section we arrive at a result equivalent to Eqn.~\ref{SEgd} via an 
alternative 
approach. We consider a thermodynamic system characterized by its
total energy, entropy, volume and energy scale $m$. We 
then make two assumptions:

\begin{enumerate}
\item The energy is an {\it extensive} function of volume and entropy.

{\it i.e.}, under a scaling of $V$ and $S$ by a factor $\lambda$,
\beq\label{extensive}
E(\lambda V,~ \lambda S)~=~\lambda~ E(V,~S)~.
\eeq

\item The total energy is parametrized by the following ansatz:
\beq\label{Eparam}
E=E_{ns}~\left(\frac{E_{ns}}{m}\right)^{\tilde g}~,
\eeq

where $E_{ns}$ stands for the energy in the case where there are no other 
scales (such as $m$) entering in the description of the system. This 
{\it no-scale} energy, by dimensional analysis, must 
be proportional to $V^{-1/d}$, {\it i.e.}, 
\beq\label{Ens}
V^{1/d}E_{ns}(V,~S)\equiv h(S)~,
\eeq
is a function of entropy only.  
\end{enumerate}
Note that in Eqn.~\ref{Eparam} we 
parametrized the dependence on the extra scale $m$ as a power law 
in analogy with Eqn.~\ref{gendisp}. 

\subsection{The scaling result}
\label{Sect:TSC}

Rescaling volume and entropy by $\lambda$ we get from Eqn.~\ref{extensive} 
and Eqn.~\ref{Ens}
\beq
E(V,~S)&=&\frac{1}{\lambda} E(\lambda V,~\lambda S)\nonumber\\
&=&\frac{h(\lambda S)}{\lambda^{1+1/d}V^{1/d}}
\left(\frac{h(\lambda S)}{m\lambda^{1/d}V^{1/d}}\right)^{\tilde g}\nonumber\\
&=&\left[\frac{\left(h(\lambda S)\right)^{\tilde g+1}}
{(\lambda S)^{(d+\tilde g+1)/d}}
\right]\frac{S^{(d+\tilde g+1)/d}}{V^{(\tilde g+1)/d}m^{\tilde g}}~,
\eeq
that is consistent only if the quantity in square brackets is independent of 
$S$. 
This implies
\beq
S(E,\tilde g,d)= S_0(\tilde g,d)
\left(\frac{E}{k_V}\right)^{\frac{d}{d+\tilde g+1}}
\left(\frac{m}{k_V}\right)^{\frac{\tilde g d}{d+\tilde g+1}}
~. \label{Sofg}
\eeq
The similarity with Eqn.~\ref{SEgd} is suggestive of a close relation to the 
statistical approach that will be explained in the next section\footnote{
The choices made above are orthogonal to those of \cite{Verlinde:2000wg}. 
There,
assumption 1 was dropped to include a {\it sub}-extensive Casimir energy 
component $E_C$ while keeping the no-scale property 
($\tilde g=0$ in assumption 2). 
This leads to the generalized Cardy formula
$S\sim V^{1/d}\sqrt{E_C(E-E_C)}~.$
We could generalize further our approach to include such a non-extensive 
component to arrive at the analog expression when the no-scale property
is lost.     
}. 

\section{Comparison of the two approaches}
\label{Sect:Compare}

The two approaches to obtain the density of states are designed to reflect 
different physical properties. The first one deals with variations of the 
energy density and pressure due to an altered dispersion relation in a way 
proper to statistical mechanics; using the representation of those quantities 
in terms of momentum distribution functions. The change in the dispersion 
relation is taken as the measure of breaking of Lorentz invariance. 
The second approach, in contrast, is centered in the response of the system 
to changes of scale, that is, in the breaking of conformality in the system. 
The results show, however a tight interplay of the two approaches. Indeed, 
if we consider a single particle state, we obtain a particular case of  
Eqn.~\ref{Eparam}, namely,
\beq\label{Esps}
{\cal E}={\cal E}_{ns}\left(\frac{{\cal E}_{ns}}{m}\right)^{\tilde g}~,
\eeq
where ${\cal E}_{ns}$ is the no-scale energy for a single particle. But, since 
we assumed the total energy to be extensive, we can sum all single particle 
states (with their corresponding density, for example, in momentum space) to 
obtain the total energy, namely,
\beq\label{sum}
E=V\frac{n}{(2\pi)^d}\int f(p) {\cal E}(p)~d^dp~,
\eeq
where ${\cal E}(p)$ should be replaced with the expression on the rhs 
of Eqn.~\ref{Esps}. 
Thus, by making the identifications 
\beq{\cal E}_{ns}(p)&\equiv& p~,\\
\tilde g&\equiv& g~,
\eeq 
we see that Eqn.~\ref{Esps} coincides with the dispersion relation 
Eqn.~\ref{gendisp} and the energy density becomes equivalent to the 
one derived in the statistical approach. 
Furthermore, setting the undetermined factor $S_0(g,d)$ in  Eqn.~\ref{Sofg} 
to $(d+2g+2)F^{\frac{g+1}{d+g+1}}/d$, makes the entropy identical to
the statistical counterpart in Eqn.~\ref{SEgd}. 

We note though that this last step is done simply by construction, and
is not a derivation. The price to pay for the simplicity and
generality of the scaling argument is that the coefficient $S_0$
remains undetermined, while its numerical value is explicit in the
statistical mechanics approach.

%%%%%%

\section{Analysis}
\label{Sect:Analysis}
We are now ready to take up our original goal: The comparison of the
field theory density of states with the Wigner semicircle.  We have
generalized the field theory density of states to include the parameter
$g$ which determines the dispersion relation according to
Eqn. \ref{gendisp}.  Since our results preserve the general form taken
in Eqn. \ref{df} for the field theory density of states, we can
continue to use the formalism reviewed in section
\ref{Sect:Approach}. 

Specifically, comparing  Eqn. \ref{SEgd} with Eqns. \ref{df} and
\ref{NofS} leads to 
\beq
b&\equiv&\frac{2g+2+d}{d}F^{\frac{g+1}{d+g+1}}
\left( \frac{m}{k_V}   \right)^{\frac{g d}{d+g+1}}~, 
\label{bgen}
\\
\alpha & \equiv &\frac{d}{d+g+1}~,
\\
\Delta _2&\approx &\frac{\alpha^2b^2}{\gamma}
\left(\frac{E_0}{k_V}\right)^{2\alpha}
\frac{\Delta E^2}{E_0^2} = \frac{\alpha^2}{\gamma}
{S(E_0)^2}\frac{\Delta E^2}{E_0^2}~.\label{d2}
\eeq
where 
\beq
&&{S}\left( {E_0,g,d} \right) =\nonumber\\ 
&&~ {\frac{{d + 2g + 2}}{d}}
{F^{\frac{{g + 1}}{{d + g + 1}}}}{\left(
  {\frac{{{E_0}}}{{{k_V}}}} \right)^{\frac{d}{{d + g + 1}}}}{\left(
  {\frac{m}{{{k_V}}}} \right)^{\frac{{gd}}{{d + g + 1}}}} 
~. \label{SofgE0}
\eeq
Recall that $\Delta_2$ measures the fractional discrepancy between
the field theory and Wigner densities of states when evaluated over an
energy range $\Delta E$ centered at energy $E_0$. The main new
ingredient here is that $\Delta_2$ now has a dependence on the dispersion
relation via the parameter $g$.  We start with an analysis of this new $g$
dependence and then place this in the context of a full analysis of
all the parameters in $\Delta_2$, the rest of which have already been
analyzed in \cite{Albrecht:2007mm} and \cite{Albrecht:2008bj}.

The smaller the value of $\Delta_2$, the better the field theory can
be approximated by the Wigner semicircle (and thus by a random
Hamiltonian).  We therefor expect values of $g$ that lead to smaller
$\Delta_2$ to be strongly favored in our scheme, because these values
will lead much more easily to a field theoretic interpretation of a random
Hamiltonian.  

\begin{figure}
\includegraphics[width=3.3in]{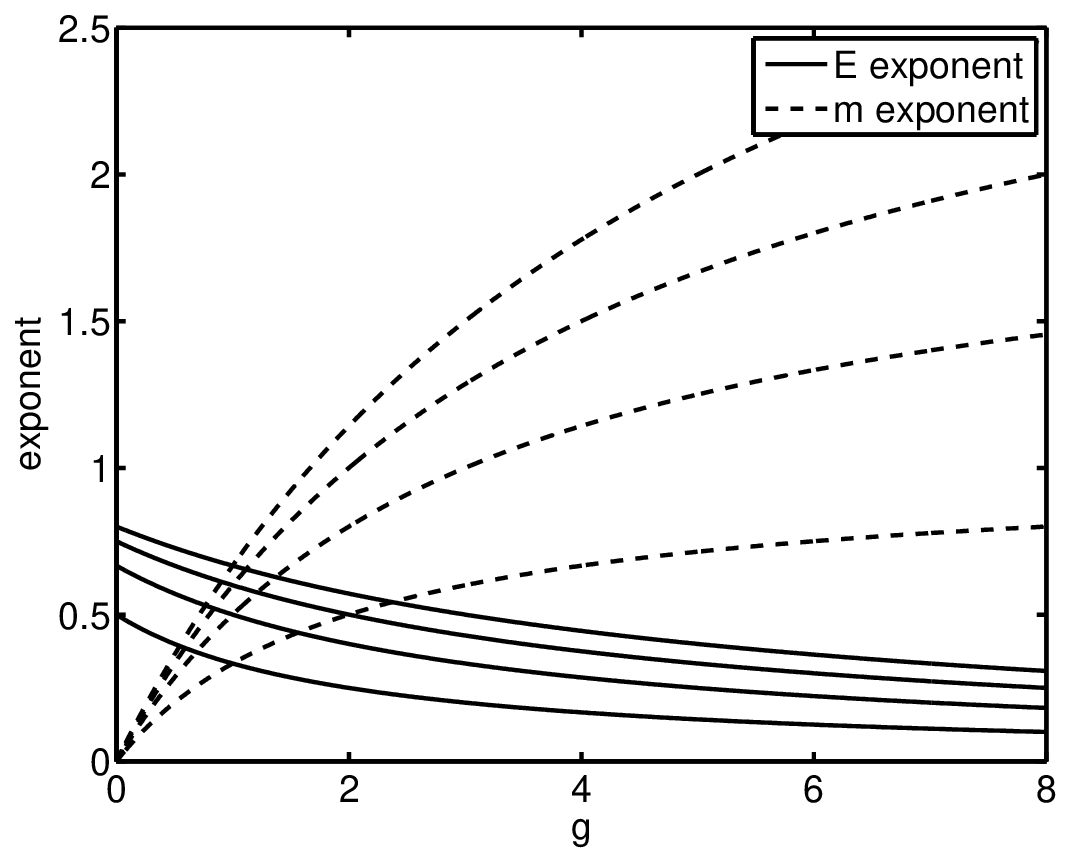}
\caption{\label{ExponentsInS}
The exponents of the $E$ factor (solid)  $m$ factor (dashed) in
Eqn. \ref{SofgE0} evaluated at $d \in \{1,2,3,4\}$ (height of the
curve increases with with increasing $d$).  Note that the $E$ exponent
decreases with increasing $g$ while the $m$ exponent grows.}
\end{figure}

\begin{figure}
\includegraphics[width=3.3in]{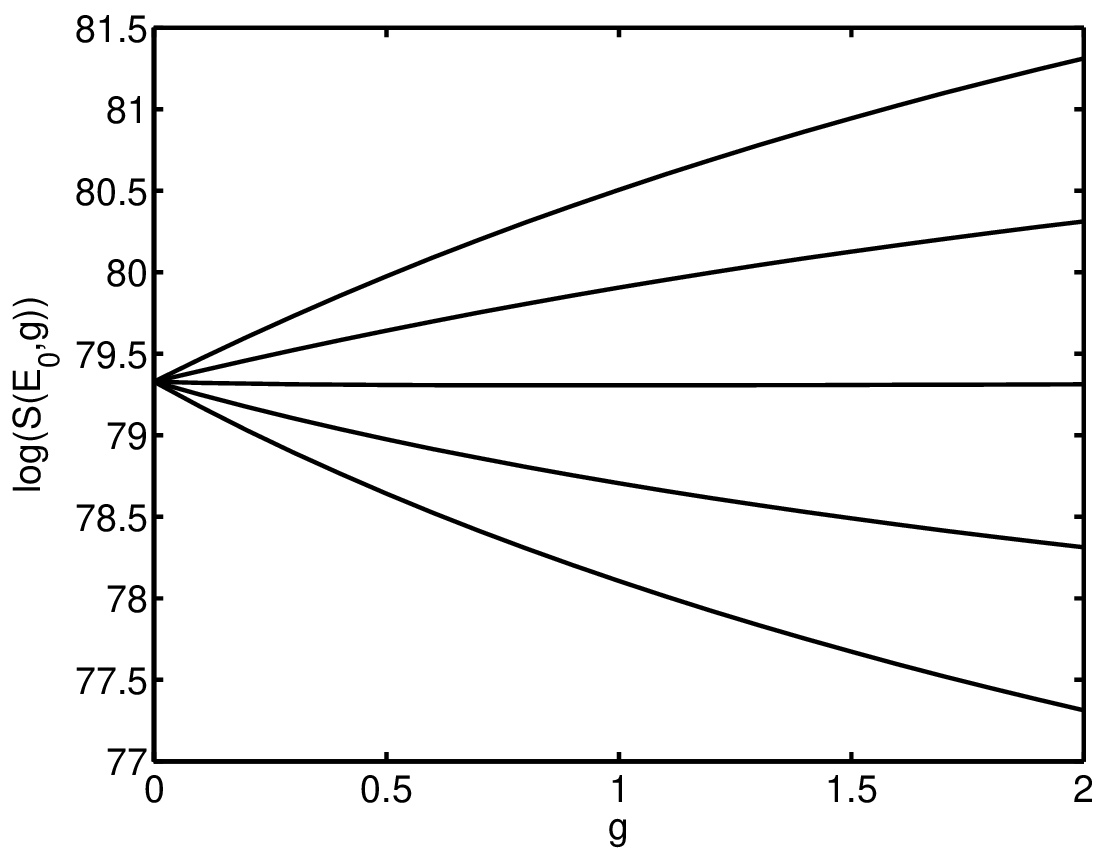}
\caption{\label{Sofgandm}
The entropy $S(E_0,g)$ for $d=3$ for 
$m/m_c = 0.01$, $0.1$, $1$, $10$ and $100$
in order of lowest to highest curves.  We
see that taking $m=m_c$ removes the $g$-dependence of $S$.  For
$m>m_c$, $S$ increases with $g$, whereas $S$ decreases for $m<m_c$  . (Here
$E_0=10^{80}GeV$ and $k_V = 10^{25}GeV$) }
\end{figure}

The dominant contribution to the $g$ dependence of
$\Delta_2$ comes from the $S(E_0)^2$ factor in Eqn.~\ref{d2}.  In turn, the
dominant contributions to the $g$ dependence of $S(E_0)$ come from the
appearance of $g$ in the exponents of $E_0$ and $m$ in Eqn.~\ref{SofgE0}.  
Figure \ref{ExponentsInS} illustrates that the exponent of the $E$
factor in Eqn. \ref{SofgE0} decreases
with increasing $g$, while the exponent of the $m$ factor grows with
$g$. 
The values of $g$ that lead to smaller values of
$\Delta_2$ thus depend on which factor
dominates. Since (as discussed near Eqn.~\ref{SEgd}) both exponentiated
factors in Eqn. \ref{SofgE0} are always greater
than unity, the question of which behavior dominates is related in a
simple way to the relative sizes of $m$, $E_0$ and $k_V$.   
A helpful quantity is
\begin{equation}
{m_c} \equiv {\left( {E_0k_V^d} \right)^{1/\left( {d + 1} \right)}}~.
\label{mcdef}
\end{equation}
Evaluating Eqn. \ref{SofgE0} at $m=m_c$ gives
\begin{equation}
S\left( {m = {m_c},g,d} \right) =  {\frac{{d + 2g + 2}}{d}}
{F^{\frac{{g + 1}}{{d + g + 1}}}}{\left(
  {\frac{{{m_c}}}{{{k_V}}}} \right)^d}~.
\label{Sofmc}
\end{equation}
Equation \ref{Sofmc} shows that when $m$ is set to the critical value
$m_c$ the two exponentiated factors in Eqn. \ref{SofgE0} can be
collected into a single factor which has no $g$ dependence, as
illustrated in Fig.~\ref{Sofgandm}. 

Figure \ref{Sofgandm} also shows that values of the dispersion
relation scale $m$ less than $m_c$ make $S$ decrease as a function
of $g$ and thus favor divergently large values of $g$,
whereas $m>m_c$ 
causes $g=0$ to be favored.  Since large $g$ corresponds to a
non-local field theory, we conclude that to make a local field
theoretic approximation to a random Hamiltonian one must choose
$m>m_c$ in the dispersion relation.  Having made that choice, the
Lorentz symmetric case, $g=0$ is favored, in that $g=0$ minimizes
$\Delta_2$ and thus can be best approximated by the random
Hamiltonian.  

\begin{figure}
\includegraphics[width=3.3in]{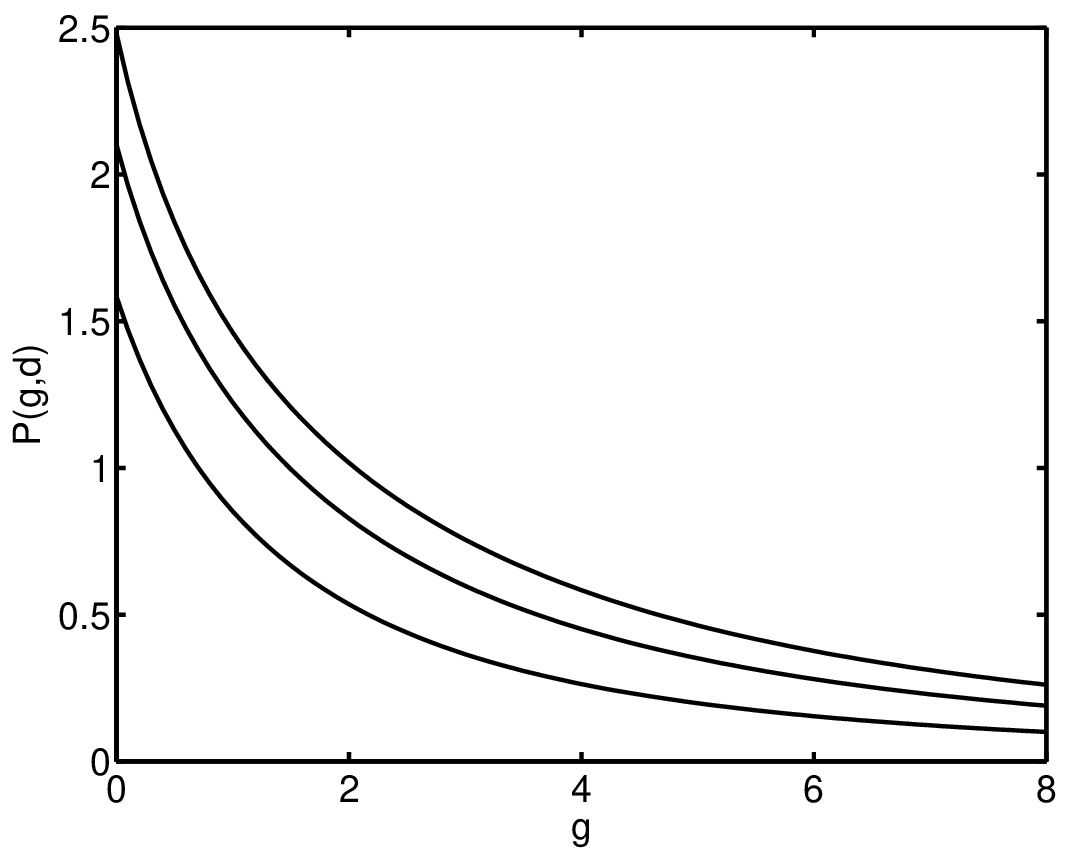}
\caption{\label{PreFactorsInD2}
The value of the prefactor $P$ (defined in Eqn. \ref{Pdef}) as a function
of $g$. The three curves show $d=2$, $d=3$ and $d=4$, lowest to
highest.  The fact that $P(g,d) = O(1)$ for a wide range of values
is why we can neglect $P$ when building intuition about the dependence
of $\Delta_2$ on $g$.}
\end{figure}

\begin{figure}
\includegraphics[width=3.3in]{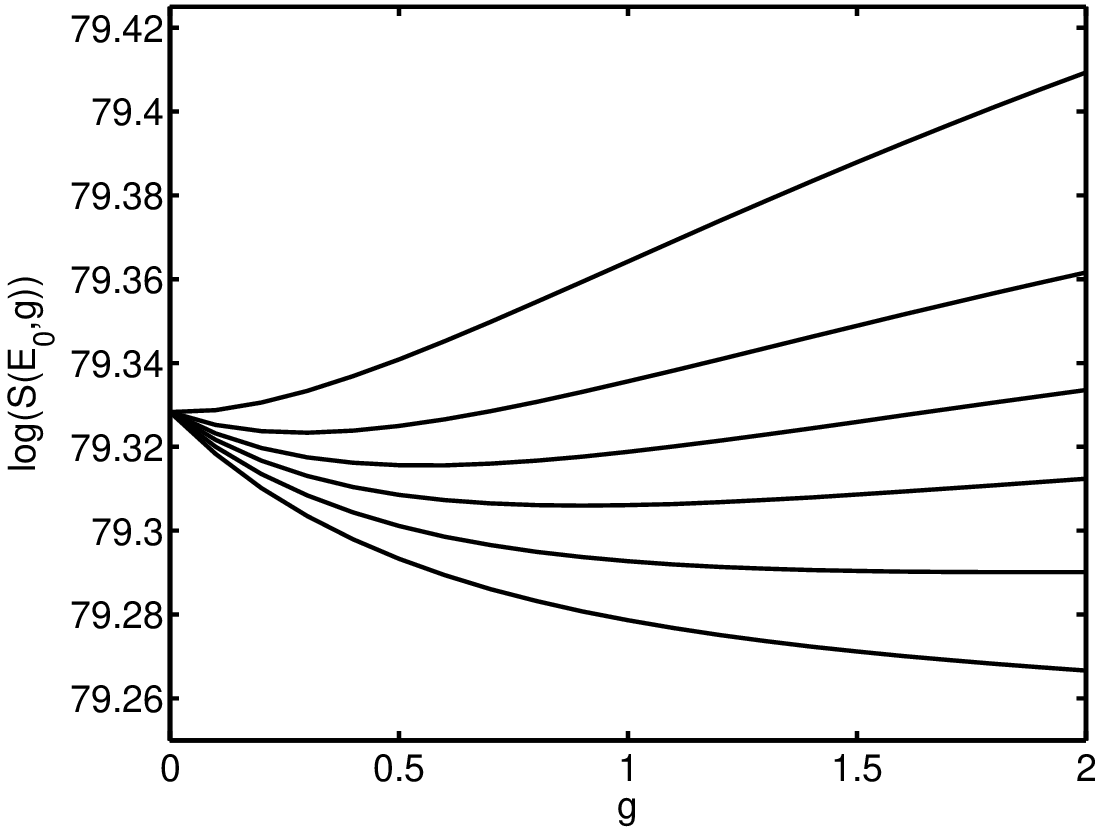}
\caption{\label{SofgandmSmall}
This figure is similar to Fig. \ref{Sofgandm} but shows values of $m$ much
closer to $m_c$ ($m/m_c = 0.9$, $0.95$, $1$, $1.05$ and $1.25$).
For $m$ so close to $m_c$ the non-trivial shape of the prefactor shown
in Fig. \ref{PreFactorsInD2} becomes visible and can lead to local
minimum somewhat away from $g=0$. That shifts the value of $m$ for
which $g=0$ is preferred to a value slightly higher than $m_c$.}
\end{figure}

For completeness we now examine the dependence of $\Delta_2$ on $g$
from contributions other than the dominant $E$ and $m$ factors in
$S$.  We define the ``prefactor'' $P$ from Eqns. \ref{d2} and \ref{SofgE0}
as
\begin{equation}
P(g,d) \equiv {d + 2g + 2 \over d}\alpha^2  F^{g+1 \over
  d+g+1}~.
\label{Pdef}
\end{equation}
which contains additional $g$-dependence not accounted for
by the exponents of the $E$ and $m$ factors in
Eqn. \ref{SofgE0}.  Figure \ref{PreFactorsInD2}
shows explicitly that
$P$ is $O(1)$ for a range of parameters as claimed.  Still, for $m$
very close to $m_c$ the shape of $P(g)$ can have an impact, as
illustrated in Fig.~\ref{SofgandmSmall}.

Since we have argued that the $g$ dependence of $\Delta_2$ drives us
to the the Lorentz symmetric case already considered in earlier work,
our analysis of the dependence of $\Delta_2$ on the other parameters
carries over unchanged.  In particular, in \cite{Albrecht:2008bj} we
considered the same $d$ dependence used here and came to the
intriguing conclusion that the
$d=3$ case put $\Delta_2$ right at the edge of values allowed by
current data. 
%% This result is illustrated in Fig. \ref{fig:del2}.  \begin{figure}
%% \includegraphics[width=3.3in]{Del2.eps}
%% \caption{\label{fig:del2}
%% $\Delta_2$ as a function of $g$ for $d=3$ (lower curve), $d=4$ and $d=5$ 
%% (upper) taking $\Delta E\sim 10 GeV$ in Eqn.~\ref{d2}.  For $g=0$ and
%%   $d=3$, $\Delta_2$ starts taking on values less than unity.}
%% \end{figure}
To achieve that one had to take $k_V \approx m_\gamma = 
10^{25}GeV$, where $m_\gamma$ is the upper bound on the photon mass.
Since $k_V$ denotes the scale where continuum field theory breaks down
(not necessarily the scale of any actual volume $V$), this seems to be a very
reasonable choice. 
Note also that in \cite{Albrecht:2008bj} we simply took $b=1$ in Eqn. \ref{df}.
Here we have derived an explicit formula for $b$ (given by
Eqn. \ref{bgen}), which indeed is $O(1)$ for the $g=0$ case. 

%% In order to perform the analysis we make the following
%% choices. We%% take %% $k_V=10^{-25}GeV$, the energy scale %% $E_0$
%% as the current%% energy of the Universe $10^{80}GeV$, and the mass
%% scale%% $m$ as a %% typical cutoff scale ($SUSY$ breaking, $GUT$ or
%% $M_P$) of order %%%% $10^{{\cal O}(10)}GeV$. Under this
%% assumptions, the behavior of %%
%% $\Delta_2$ is depicted in Fig.~\ref{fig:del2}. 

\begin{figure}
\includegraphics[width=3.3in]{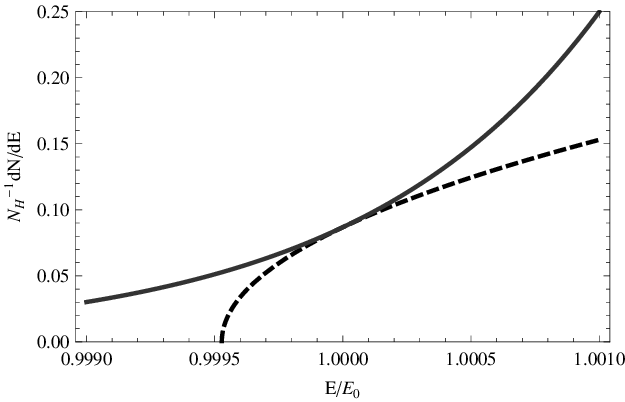}
\includegraphics[width=3.3in]{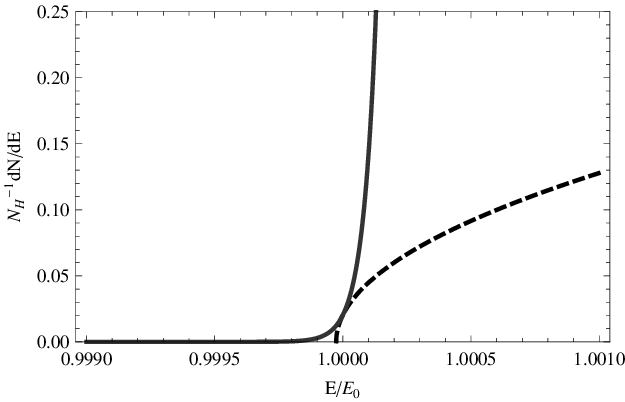}
\includegraphics[width=3.3in]{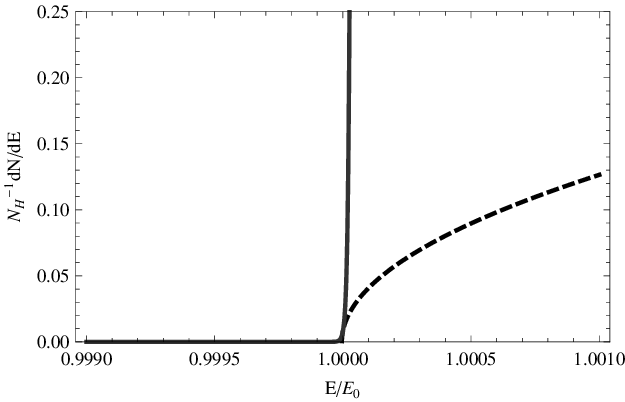}
\caption{\label{fig:dndeg}
$N_H^{-1}dN/dE$ for random Hamiltonians (dashed lines) and field theory 
(solid lines)  matching the zeroth and 
first order terms in an expansion around $E_0$ as a function of $E/E_0$ 
for $g=0$ (upper panel), 
$g=1$ (center) and $g=2$ (lower). The Lorentz symmetric ($g=0$) case
gives the broadest overlap, illustrating the main conclusion of this paper.  
The plots correspond to $d=3$, 
$\beta=2$, $\gamma=1/2$, $k=10{-3}E_M$,
$E_0=8E_M$ and $m=E_M$. The parameters $N_H$ and $E_S$ were solved for as
explained in section \ref{Sect:Approach}. 
}
\end{figure}

As already discussed, $\Delta_2$ quantifies the degree to which the
density of states curves for a field theory and a random Hamiltonian
can over lap over an energy range $\Delta E$.  Figure \ref{fig:dndeg}
gives an explicit illustration of what is going on, where one can see
how the two curves match more closely for the $g=0$ case.  In this
figure, some differences between the curves from one panel to the next
reflect the different impacts of enforcing equality at zeroth
and first order (Eqns. \ref{m1} and \ref{m2}) for different values of $g$.

\section{Discussion and Conclusions}
\label{Sect:Conclusions}
We have calculated the density of states for a free field theory with
an arbitrary power law dispersion relation.  This allowed us to extend
our earlier work examining the degree to which a random Hamiltonian
could be approximated by a local field theory.  Our conclusion is that
choosing the Lorentz invariant ${\cal E}(p) = p$ dispersion relation
reduces the discrepancy between the energy eigenvalue spectrum of a
local field theory and that of the random Hamiltonian, as
compared with the other dispersion relations considered.  In this
sense, Lorentz symmetry is favored by our analysis.  This is the
concrete result from our work.

As discussed in \cite{Albrecht:2007mm,Albrecht:2008bj}, quite a few
steps still need to be taken to check the viability of the ``random
Hamiltonian'' picture. The work presented here offers additional hints about
specific areas for further investigation.  
As emphasized in the introduction, by focusing on the dispersion
relation we are unable to directly draw conclusions about the
full Lorentz symmetry of the theory, although it is possible that
preferences (for or against) could emerge from a deeper analysis that
connects the eigenstates to observable particles.   In any case,  
imposing Lorentz invariance on the field theory will be a quite
generic way to ensure the preferred dispersion 
relation.  One then would expect massless spin-1 and spin-2
representation could turn up in the interpretation of a random
Hamiltonian as a field theory. According to general arguments
\cite{Feynman:1996kb,Weinberg:1964ev,Weinberg:1964cn,Weinberg:1965nx,
Weinberg:1965rz,Deser:1969wk,Weinberg:1980kq} 
these fields would
lead to gauge forces and general relativity respectively.  If we assume that 
the random Hamiltonian is the
correct underlying theory, this picture should provide a systematic
and hopefully testable way in which local field theory, gauge theory and
gravity will break down.  

Looking more specifically at the technical results in this paper, we
chose $m>m_c$ (with $m_c$ defined in Eqn. \ref{mcdef}) to avoid favoring 
arbitrarily large powers of momentum $p$ in the dispersion relation 
\begin{equation}
  {\cal E} \left( p \right) = p{\left( {\frac{p}{m}} \right)^g}
\label{gendisp2}
\end{equation}
which would give a nonlocal
field theory.  This led to a preference for $g=0$ which seems to 
eliminate $m$ from the picture altogether.  However, $m_c$ may show up
in some higher order way:  The two values considered in
\cite{Albrecht:2007mm} and \cite{Albrecht:2008bj} for $k_V$ are $k_V =
(V_U)^{1/d}$ (where $V_U$ is the Hubble volume) and
$k_V = m_\gamma$. Continuing to use $E_0=\rho_U V_U=10^{80}GeV$ as we
have here and in our earlier work, 
these give $m_c = \rho^{1/d}$ and $m_c \approx 1GeV$
respectively. Both are interesting scales for physics in our
world. Note the scale $\rho^{1/d}$ is associated with both the dark
energy and the neutrino mass\footnote{
We found in \cite{Albrecht:2008bj} that the value $k_V = (V_U)^{1/d}$
only passed our energy spectrum test when an alternate form of the
density of states was employed. This form was based on
\cite{Verlinde:2000wg} (as we mentioned briefly in section \ref{Sect:TSC}).
One would need to extend results of
\cite{Verlinde:2000wg} to include generalized dispersion relations to fully
explore the $k_V = (V_U)^{1/d}$ case.
}.  
One way $m_c$ could enter is through the fact (illustrated in Fig.
\ref{SofgandmSmall}) that local Lorentz symmetry breaking theories are
preferred for $m$ very close to $m_c$. We find it intriguing that
other authors have associated neutrino mass with
Lorentz-violation\cite{Cohen:2006ir}, while our scheme appears to 
associate Lorentz-violation with the neutrino mass scale. 

The idea that physics at its most fundamental is simply described by a random
Hamiltonian is a curious idea which we motivated in previous work on the
clock ambiguity.  There is much still to be explored before such an
idea could look seriously viable.  Still, we find it interesting that
this idea has passed as many tests as it has so far.  In particular,
this paper shows that the random Hamiltonian picture predicts
that we would interpret the physical world using theories with Lorentz 
symmetric dispersion relations.  We find this result quite remarkable.

% If you have acknowledgments, this puts in the proper section head.
\begin{acknowledgments}
We are grateful to Cliff Burgess, Steve Carlip, Teresa Hamill, Nemanja Kaloper, Don Page and
Dan Phillips for helpful discussions. AA thanks the
Perimeter Institute where a number of these discussions took
place. AI thanks the University of California at Davis for 
hospitality while this work was completed. This work was supported in part by 
DOE Grant DE-FG03-91ER40674 and the Humboldt-Foundation.   
\end{acknowledgments}

% Create the reference section using BibTeX:
%\bibliographystyle{apsrev}
\bibliography{Clock}

\end{document}